\documentclass[revtex4]{article}
\usepackage[affil-it]{authblk}
\usepackage{graphicx}
\usepackage{fancyhdr}
\usepackage{multicol}
\usepackage{amsfonts}
\usepackage{framed,color}
\definecolor{shadecolor}{rgb}{1,0.8,0.3}
\usepackage{fancybox}
\usepackage{mdframed}
\usepackage[tikz]{bclogo}

\begin{document}
\title{A holographic bound on the total number of computations in the visible Universe} 
\author{Maurice H.P.M. van Putten}
\affil{Department of Astronomy and Space Science, Sejong University, 98 Gunja-Dong Gwangin-gu, Seoul 143-747, Korea}
\maketitle
\begin{abstract}
Information $I$ in holographic imaging of massive particles by star-like screens is shown to represent the probability 
of detection based on their propagator. Results are derived for screens in the shape of a plane, cube and sphere from 
unitarity in the exponentially small transition probability for a detection outside. 
We derive $I=2\pi \Delta\varphi$ in $\log2$ bits for the imaging of a particle by a spherical screen
at a relative de Broglie phase $\Delta\varphi$. Encoding mass, charge, angular momentum or 
radiation requires at minimum four bits. Minimal screens at maximal information density hereby recover Reissner-Nordstr\"om and extremal Kerr black holes. Applied to the visible Universe, the Hubble flow of galaxies through the cosmological event horizon leaves $10^{121}$ computations in the future.
\end{abstract}

 
\section{Introduction}

Confronted with quantum mechanics, event horizons of black holes in general relativity are naturally associated
with entropy and temperature consistent with thermodynamics \cite{haw77}. More generally, the entropy 
of matter and fields in three-space is bounded by the enclosed mass-energy times the radius of an enclosing sphere \cite{bek81}.

The holographic principle proposes that matter and fields are to be viewed as images produced by two-dimensional 
time-like screens \cite{thoo93}, which naturally includes gravitational focusing \cite{sus95}. 
Closely related to surface area is the action integral, which suggests that the holographic principle is 
related to canonical quantizations of matter and fields with potentially interesting cosmological applications 
\cite{llo02}. However, a general computational framework remains elusive, in part, due to our lack of understanding
of the nature of computation in a holographic representation of the universe. Some detailed entropy calculations
are only recently being explored by computations in space-time near event horizons based on 
string theory \cite{han09,han13,hya13}. 

Here, we set out to calculate the information required for the encoding of massive particles
from their quantum mechanical propagator. A holographic screen creates a partition of three space into
a region inside and outside. The propagator defines transition probabilities for a particle prepared inside
to be detected outside. A holographic encoding expresses the logarithm of these probabilities in surface
elements on the screen. This suggests a framework for some concrete calculations based directly from
propagators and holography by two-surfaces. Any discrete representation of information on the screen, e.g., 
in bits, hereby leads to a quantization of phase space within. 

Our approach is complementary to entanglement entropy \cite{bom86,fro97a,fro97b,isr03}
in the division of the quantum system introduced by the screen \cite{mot10,shi11} .

In a binary encoding in finite Planck-sized surface elements, the size of holographic screens is naturally bounded 
below by the Schwarzschild radius of the enclosed particle. Following a detailed discussion on
the partition of energies, this bound can be generalized to charged particles.
In the application to cosmology, the cosmological event horizon introduces an
upper bound on the size of a holographic screen. The result will be found to have implications for
the problem of computation in cosmology. 

Following notation, we describe in \S2 information in cuts by planes and in \S3 by cubic screens. \S4 discusses 
focusing in wave propagation in general relativity and \S5 information in spherical screens.  In \S6, minimal binary 
encoding is discussed. Following a covariant formulation of the results of \S5 in \S7, the four-bit encoding is shown to
derive Reissner-Nordstr\"om and extremal Kerr black hole solutions. In our application to cosmology, we encounter
a remarkable efficiency of computation in Nature. The results indicate a novel bound on the
total number of computations in the future of the visible Universe in response to the Hubble flow of galaxies
through the cosmological event horizon (\S8). We summarize some of our findings in \S9. 

\section{Notation and preliminaries}

We start with a topological aspect of oriented two-surfaces of partitioning space into two disjoint regions. 
A plane of infinite extent introduces space to the left with a complement to the right. More generally, a 
closed two-surface introduces regions inside and outside according to the Jordan-Brouwer separation 
theorem \cite{ale22,gui10}. In the presence of matter, such partition of the domain of a particle wave function
introduces probabilities for the mutually exclusive alternatives of particle detections in either one of the two regions. 
We shall refer to such partition (of the domain) of a wave function by a two-dimensional
screen as a ``cut." For a holographic screen, this raises the question how much information is associated 
with the probability of detecting a particle inside (or to the left) and the encoding thereof in elementary memory 
units on its surface.

To be specific, consider the de Broglie matter wave of a particle of mass $m$.
The wave number $k=mc/\hbar$ reduces to $k=m$ in natural units with the velocity of light
$c=1$ and the reduced Planck constant $\hbar=1$. Given Newton's constant $G$, we further have the Planck
length $l_p=\sqrt{G\hbar/c^3}$, whereby $k\l_p^2 = R_g$ is the associated gravitational radius. Putting $G=1$,
we have $m=R_g$ and a Schwarzschild radius $R_S=2R_g$.

Given the action integral $S=m\int_A^B ds$ over a path with end points $A$ and $B$,
we have the transition amplitude
\begin{eqnarray}
<B|A> = e^{iS},~~S=[\phi]_A^B \chi(A,B),
\label{EQN_AB}
\end{eqnarray}
where $[\varphi]_A^B=\phi(B)-\phi(A)$ denotes the difference in phase in the de Broglie wave function
at $A$ and $B$. $\chi(A,B)=\{1,i\}$ represents the local causal structure of a light cone, where the two
values are associated with time like or, respectively, space like separations between $A$ and $B$.
In case of a time like separation, the amplitude $<B|A>$ is described by Feynman's phase factor $e^{iS}$, 
whereas in case of a space like separation, $<B|A>$ is described by the tunneling amplitude $e^{-S}$
\cite{ton07}.

Consider a particle which is prepared at $A$ to the left of a two-dimensional screen $\Sigma$ of infinite extent. 
The cut of the wave
function by $\Sigma$ defined above has one degree of freedom given by its distance to $A$. Here, the distance is expressed by the transition amplitude for the particle to be found to the right of $\Sigma$ in a subsequent measurement. By covariance of (\ref{EQN_AB}), the resulting transition probabilities are Lorentz invariant. Integrating out momenta 
of the particle at $A$ and $B$, the result is a propagator (e.g. \cite{ton07}). Whenever the distance of $\Sigma$ to $A$ 
is macroscopic relative to the de Broglie wave length $\lambda=2\pi/k$, the transition probability will be exponentially 
small since $\chi(A,B)=i$ for space like separations. Let $P_+$ denote the probability of finding the particle to the left.
Then the probability $P_-$ for a detection to the right in a subsequent measurement satisfies
\begin{eqnarray}
P_- + P_+ \equiv 1
\label{EQN_UN}
\end{eqnarray}
by unitarity. It defines the information in a cut, i.e., the Shannon information $-\log P_+$ is required to satisfy
(\ref{EQN_UN}) {\em exactly}, i.e., $P_\pm$ must obtain by exact arithmetic on a finite state machine. 

We set out to describe the relation between information of a cut and surface area associated with curvature and 
information encoding. The result illustrates holographic imaging by focusing of null trajectories as described
by the Raychaudhuri equation, wherein particles appear by interference of matter waves. Our discussion
is hereby different from existing discussions on entropic considerations, in relying on (\ref{EQN_AB}-\ref{EQN_UN}) 
with no reference to temperature or thermodynamics.

\section{Information in a plane and a cubic screen}

Following (\ref{EQN_AB}), a particle prepared at a location ${\bf r}_1$ at $A$ has a probability $P_-$ to be detected 
at ${\bf r}_2$ at $B$ in a subsequent measurement, satisfying \cite{ton07}
\begin{eqnarray}
p(s) = \left<0|\phi({\bf r}_2)^\dagger\phi({\bf r}_1)|0\right>\simeq e^{-ks},
\label{EQN_p1}
\end{eqnarray}
where $s=|{\bf r}_2-{\bf r}_1|$ is the space like separation between $A$ and $B$. We here tacitly integrate out the phase space of momenta at ${\bf r}_\pm$ as well as time between measurements. 

If the particle is prepared at one side of a screen at a distance $s$, it leaves a small but nevertheless finite probability $P_-$ to be subsequently detected at the other side (Fig. 1). This probability is exponentially small whenever $s$ is larger than the de Broglie wave length of the particle. An elementary calculation shows
\begin{eqnarray}
P_-(s) \simeq \int_{H^-} e^{-2kr} d^3x \simeq \frac{\pi s}{2k^2}e^{-2ks}.
\label{EQN_Po}
\end{eqnarray}
It varies with the distance $s$ from being exponentially close to $1$, to $\frac{1}{2}$ on the 
screen at $s=0$, to exponentially small across. By (\ref{EQN_UN}), the cut of a wave function created by a screen 
is hereby parameterized by a probability. The phase difference $[\varphi]_A^B=ks$ across a space like separation 
$s$ is a covariant factor in the action integral (\ref{EQN_AB}). Subject to a Lorentz transformation, the distance 
interval $s$ may contract by a Lorentz factor $\Gamma$, but the product of mass $m=\Gamma m_0$ and 
$s/\Gamma$ is constant.
\begin{figure}
\centerline{
\includegraphics[width=40mm,height=40mm]{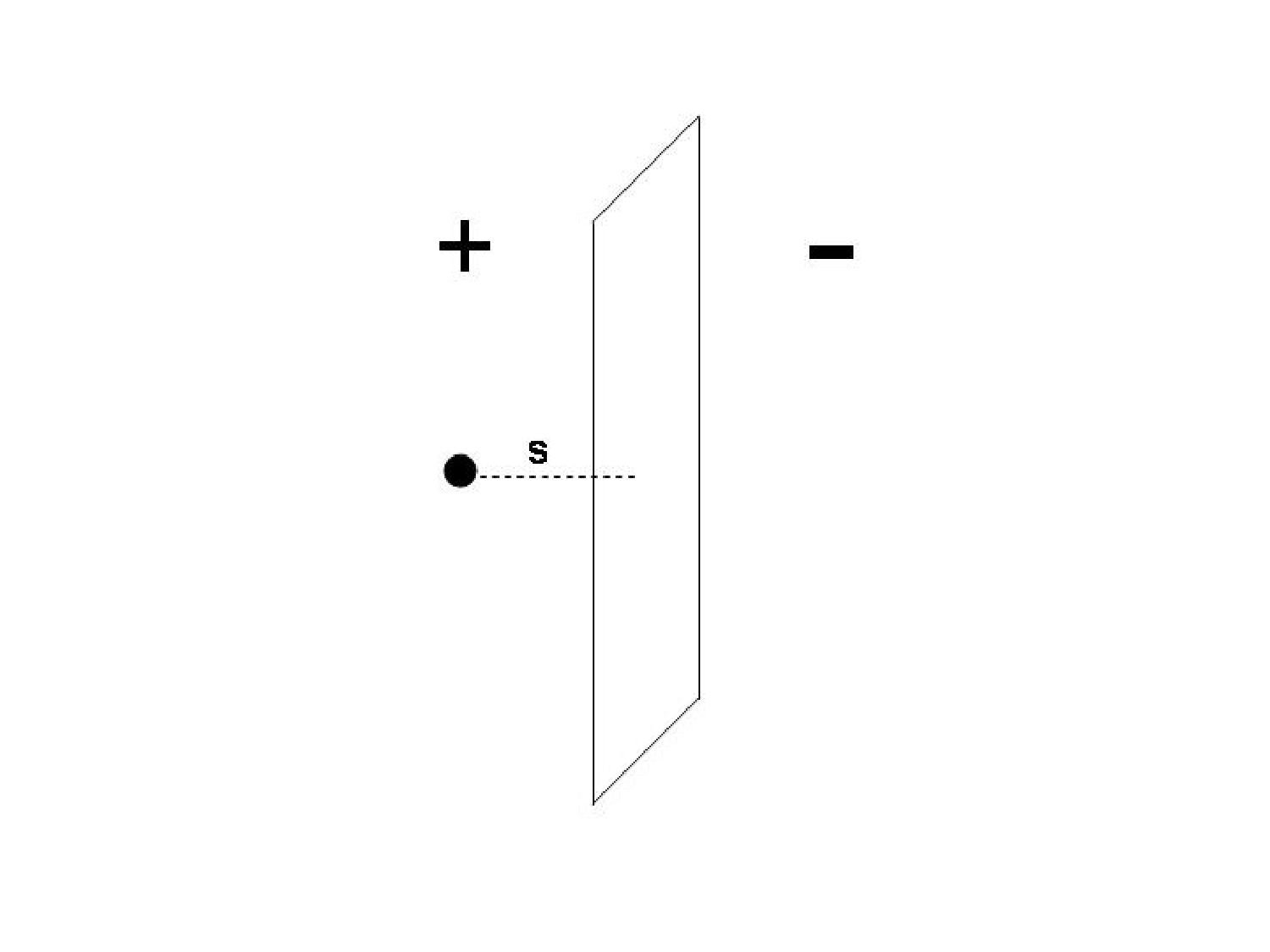}\includegraphics[width=40mm,height=40mm]{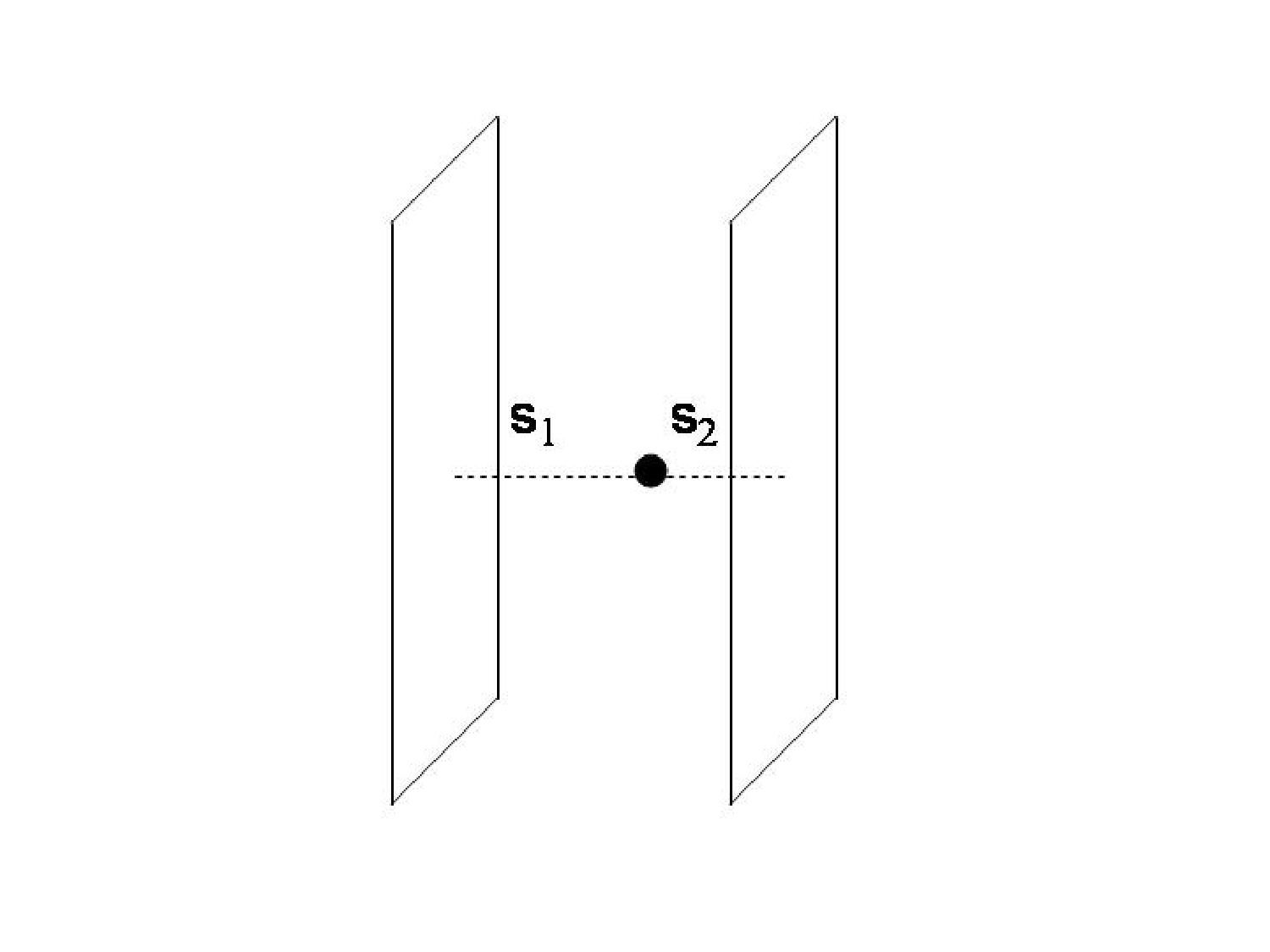}
\includegraphics[width=40mm,height=40mm]{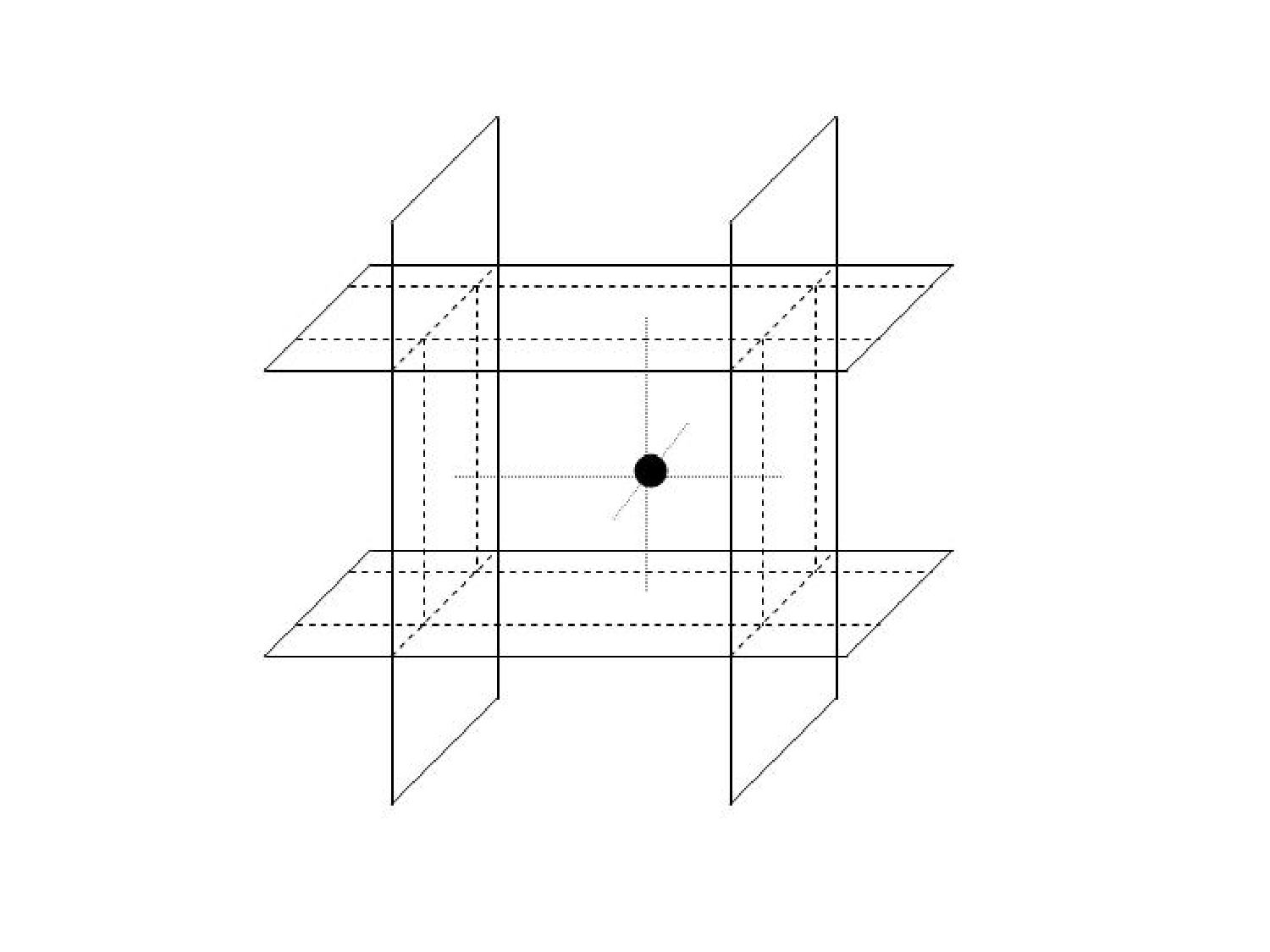}}
\caption{($Left.$) A flat screen of infinite extent partitions space in region $H^\pm$ to the left $(+)$ and to the right $(-)$, where
the probabilities of finding a particle satisfies unitarity. Prepared in ($+$) at  distance $s$ from the screen, 
$I=-\log P_+ \simeq 2ks$ is the information in the cut of the wave function. The information in two independent plane
parallel cuts at distances $s_1$ and $s_2$, $s=s_1+s_2$, to $m$ satisfies $I=2ks$ ({\em middle}) and, for a cube $(right)$, $I=6ks$, where dashed lines refer to the intersection of the third pair of parallel planes.}
\label{fig:F1}
\end{figure}

The Shannon information $I$ in (\ref{EQN_Po}) is $\log 2$ times the number of bits required to accurately resolve its numerical value. At macroscopic separations relative to the de Broglie wave length of the particle, $P_-$ is exponentially small. In this limit,
the number of bits required scales linearly with the separation $s$. To be precise, $I=-\log P_-$ satisfies
\begin{eqnarray}
I(s) \simeq   2ks,
\label{EQN_I1}
\end{eqnarray}
where we neglect higher order logarithmic terms. Two {\em independent} cuts by additional parallel planes at positions given by separations $s_1$ and $s_2$ to $m$ carry information (\ref{EQN_I1}) with $s=s_1+s_2$ (Fig. \ref{fig:F1}). 

If $I(s)$ is encoded in units of one bit, a 1-bit uncertainty in (\ref{EQN_I1}), $1=2k\Delta s$, satisfies the uncertainty relation 
\begin{eqnarray}
\Delta p \Delta s = \frac{1}{2}\hbar, 
\label{EQN_HE}
\end{eqnarray}
where $\Delta p = mc$. In a holographic picture, therefore, the uncertainty relation (\ref{EQN_HE}) in an image follows from the
quantization of distance information in {bits}.

Extending the above, three pairs of plane parallel surfaces about $m$ forms a cube of side length $l$. Its six faces introduce six (independent) cuts of the wave function of $m$, giving a total information 
\begin{eqnarray}
I= 6 k l
\label{EQN_Ic}
\end{eqnarray}
on a total surface area of $6 l^2$. We will elaborate on this further in \S6 below. 

The binary alternatives of detecting a particle inside or outside a screen is naturally encoded in bits. In this event, (\ref{EQN_I1}) refers to 
$\log 2$ times the number of bits. However, in considering these alternative outcomes of a measurement, the presence of 
one particle is a prior. 
The true number of states in the elementary memory units may therefore be larger than two, e.g., three in
$\{-1,0,1\}$ with 0 in reference to the absence of a particle and $\{-1,1\}$ otherwise. Some support for three-state memory
units can be inferred from the spectrum of eigenmodes of black hole event horizons \cite{hod98,dre03}.

\section{Einstein area and lensing}

We next recall some basic elements from general relativity associated with massive objects.

\subsection{Einstein area}

The geometric aspect of curvature is manifest in the focusing of null trajectories. 
In what follows, we shall work in the limit of time symmetry, defined by the vanishing of the extrinsic curvature tensor in a choice of foliation in Cauchy surfaces of constant coordinate time $t$ with three metric $h_{ij}$ and associated covariant
derivative $D_i$. 

Consider the divergence $\theta=D_is^i$, where $s^i$ is the unit normal along the projection of a congruence 
of null-vectors $k^a$ ($k^2=0$) onto a Cauchy surface. By Gauss's law, the surface area of a de Broglie wave front emanating from a point particle satisfies 
\begin{eqnarray}
A_f=\int_V\theta\sqrt{h}d^3x,
\end{eqnarray}
where $h_{ij}$ denotes the three-metric of surfaces of constant time at infinity. If $\tau$ denotes the time measured 
by a local static observer, we have
\begin{eqnarray}
A_f = \int_V \left( \theta_0 + \int_0^{\tau(x^i)} \left[ \frac{d\theta}{d\tau} d\tau \right] \right) \sqrt{h} d^3 x = A- A_E,
\label{EQN_Af}
\end{eqnarray}
where $A$ is the surface area of the wave front in the limit of Euclidean space and  
\begin{eqnarray}
A_E = - \int_D \frac{d\theta}{d\tau} \sqrt{h} N d^4 x
\label{EQN_AE}
\end{eqnarray}
is the Einstein area. Here, $N$ denotes the lapse function, whereby $d\tau = N dt$, and $D$ is finite four-volume 
associated with an interval of time $\Delta t$ as measured at infinity. 
The Einstein area expresses the {\em regression} in the wave front due to focusing whenever
${d\theta}/{d\tau} < 0.$ Area regression (\ref{EQN_Af}) can be expressed in terms of focusing, 
transforming a spherical opening angle $d\Omega$ in the limit of
flat space into 
\begin{eqnarray}
d\Omega^\prime = \left( 1 - \frac{A_E}{A}\right) d\Omega.
\label{EQN_fom}
\end{eqnarray}

Let $\tau$ denote the affine parameter of the null-trajectories. Then $u_ak^a = -1$ is satisfied by the velocity four-vector
$u^b$ of a local static observer. According to the linearized Raychaudhuri equation, focusing is expressed in terms 
of the Ricci tensor $R_{ab}$,
\begin{eqnarray}
\frac{d\theta}{d\tau} = - R_{ab}k^ak^b.
\end{eqnarray}
General relativity postulates $R_{ab}k^ak^b=8\pi \rho$ in the presence of matter with density $\rho$ with three-volume integral $m$. If $m$ is at a distance $s$ from a closed surface $\Sigma$, then \cite{van12}
\begin{eqnarray}
A_f = 4\pi r^2 - 8\pi ms \rightarrow 4\pi r^2 - 8\pi mr = A\left(1-\frac{2m}{r}\right),
\label{EQN_Afc}
\end{eqnarray}
where the limit follows when $\Sigma$ is a sphere of radius $r$ around $m$ at its center. The Einstein area
\begin{eqnarray}
A_E=8\pi mr
\label{EQN_AE1} 
\end{eqnarray}
(four times the circumference times mass) represents a wave front regression in the area $A=4\pi r^2$ 
by a factor $1 -2m/r$. With maximal regression occurring for $A_f=0$, we have the general inequality
\begin{eqnarray}
A_E \le 4\pi r^2.
\label{EQN_AE2}
\end{eqnarray}

Upon identifying a wave front with a holographic screen, the Einstein area is representative for the 
information encoded in the process of holographic imaging of the enclosed mass. 

\subsection{Gravitational lensing}

Regression (\ref{EQN_Af}) attributed to focusing (\ref{EQN_fom}) reflects, equivalently, an increase in poloidal surface 
area by $A_E/4$ for a given wave front area $A_f$. In weak gravity, the trajectory of a null-geodesic appears in the
approximation of flat space with an angular deficiency $\theta_e$ about a vertex at $m$, satisfying
\begin{eqnarray}
\frac{1}{2}\theta_e r^2 = \frac{1}{4} A_E.
\label{EQN_h1}
\end{eqnarray}
A photon in the poloidal plane hereby deflects according to
\begin{eqnarray}
\frac{d\theta_e}{d\theta} = \frac{\theta_e}{2\pi } = \frac{A_E}{A},
\label{EQN_h2}
\end{eqnarray}
giving the familiar focusing angle
\begin{eqnarray}
\varphi_E = 2m\int_{0}^\pi \frac{d\theta}{r} = \frac{4m}{b}
\label{EQN_b}
\end{eqnarray}
of general relativity, where $b=r\cos\theta$ denotes the impact parameter of light passing by $m$. 
Wave front regression by the Einstein area $A_E$ in (\ref{EQN_AE1}) is inextricably linked to lensing. 

\section{Information in spherical screens}

We next turn to the problem of holography by screens of spherical shape. To derive this, we first recall some
principle facts in radiation, to establish a common ground between optical holography and holography by de 
Broglie matter waves. 

\subsection{Optical holography} 

In optical holography, the information projected out from a surface element is in proportion to its surface area $\Delta A$ and the spherical opening angle $\Delta \Omega$ subtended along a line of sight. The information projected out hereby satisfies the
same scaling as radiation intensity in electromagnetic radiation \cite{ryb79}. 

In particular, we recall the specific radiation intensity, $i$, given by the radiation per unit surface area and unit opening angle, which is conserved along a given line of sight. This property follows directly from consideration of light passing through two surface elements normal to a light ray. Let $A_k$ $(k=1,2)$ denote their surface area and $d$ their separation distance. 
Associated with the bundle of light rays passing through both surface elements are two spherical opening angles $\Omega_k$, satisfying
\begin{eqnarray}
A_2=d^2\,\Omega_1,~~A_1=d^2\,\Omega_2.
\label{EQN_L12a}
\end{eqnarray}
Since the photon flux $L_k=i_kA_k\Omega_k$ passing through each surface element is conserved,
\begin{eqnarray}
i_1 A_1 \Omega_1 = i_2 A_2 \Omega_2,
\label{EQN_L12b}
\end{eqnarray}
(\ref{EQN_L12a}-\ref{EQN_L12b}) combined implies the classical result
\begin{eqnarray}
i_1=i_2.
\label{EQN_L12c}
\end{eqnarray}

Next, consider a point source described by a specific intensity $i$ as a function of direction, e.g., 
the angles ${\bf \xi}=(\theta,\phi)$ in a spherical coordinate system $(r,\theta,\phi)$. Since a point source is a singular
limit, we may define $i$ by introducing a fiducial (small) surface area $A_1=A_s$ of the source, such that
\begin{eqnarray}
L=A_s \int_{{\bf \xi}\epsilon S^2} i\,d\xi_1d\xi_2
\label{EQN_L12d}
\end{eqnarray}
is the total luminosity, where integration is over the unit 2-sphere $S^2$. Since $i$ does not depend on $r$, we have,
\begin{eqnarray}
L=A_s \int_{x\epsilon S} i\,d\Omega_x 
\label{EQN_L12d}
\end{eqnarray}
for integration over a sphere $S$ of radius $r$ centered about the point source. 

The integral on the right hand-side of (\ref{EQN_L12d}) is invariant for any {\em star-shaped} surface $C$, possessing 
 a 1-1 mapping to $S$ by rays emanating from the origin. In particular, a cube is a star-shaped surface relative to its 
geometric center. 

We are at liberty to attribute light passing through to $C$ to a specific intensity of $C$ itself. In this event, 
$i^\prime = (A_s/A)i$, where $A$ is the total surface area of $C$, whereby
\begin{eqnarray}
L =  {A}\int_{x\epsilon C} i^\prime(x) d\Omega_x
\label{EQN_L12e}
\end{eqnarray}
for any star-shaped surface $C$. 

In optical holography, information is projected out by interference between light rays. The information on a screen in 
optical holography hereby scales with the number of light rays passing through. The information density, i.e., information 
per unit surface area and unit spherical opening angle, hereby satisfies the same conservation law (\ref{EQN_L12c}) as 
specific intensity of light with an associated total information (\ref{EQN_L12e}). In what follows, therefore, we shall
identify information $I$ in holographic encoding with $L$ in (\ref{EQN_L12e}) with the associated specific information
density
\begin{eqnarray}
i = \frac{L}{4\pi A} = \frac{1}{4\pi } \int_{x\epsilon C} i^\prime(x) d\Omega_x.
\label{EQN_L12f}
\end{eqnarray}

\subsection{Holography by de Broglie waves}

Imaging or, more generally, pattern formation, by wave interference is a common principle of wave motion, in electromagnetic 
and matter waves alike. We next consider interference in de Broglie matter waves similarly to optical 
holography.

In general relativity, parallel transport of geodesic deviation vectors along null-rays may be viewed as propagation of
information in (transverse) 2-surfaces. It represents the imprint of intervening matter, more commonly studied as gravitational 
lensing of light passing on to an observer from a background source. The information in geodesic deviation scales 
with the observer's solid angle to the image plane, in exact analogy to the specific radiation intensity in optics discussed above.

Following (\ref{EQN_Ic}), we consider cube with a mass $m$ at its center, formed by six faces each 
at a distance $s=l/2$. With a total information $I=6kl$ on a total surface area $A=6 l^2$, we have a specific information 
intensity
\begin{eqnarray}
i^\prime = \frac{k}{8\pi s}. 
\label{EQN_Ic2}
\end{eqnarray}
The cube being star-shaped, the same (\ref{EQN_Ic2}) holds for a sphere $S$ of radius $s=l/2$ within the cube,
where $s$ is much greater than the Schwarzschild radius of $m$. By (\ref{EQN_L12e}-\ref{EQN_L12f}), the total information on $S$ is
\begin{eqnarray}
I= 4\pi s^2 \int_\Omega \int_{x\epsilon S} i^\prime d\Omega = 2\pi ks.
\label{EQN_I4}
\end{eqnarray}
In passing from a cube to a sphere, we note that (\ref{EQN_I4}) is $\pi/6$, i.e., about one-half of $I=12 ks$ in (\ref{EQN_Ic}) due
to the discrepant surface areas of $C$ and $S$.

Upon multiplication of $k$ with $l_p^2$, where $l_p^2=1$ in natural units with $G=1$, (\ref{EQN_I4}) may be expressed 
in an equivalent surface area
\begin{eqnarray}
A_I= 2\pi ms
\label{EQN_I4b}
\end{eqnarray}
containing the information in holographic imaging. 

\section{Minimal binary encoding}

In the present approach, holographic encoding is that of the binary alternatives of particle detections inside or 
outside of a closed surface. This gives a natural starting point for encoding in bits consistent with the 
uncertainty principle (\ref{EQN_HE}).  On an orientable two-surface, each memory cell is oriented. This allows 
each memory state to refer a particle facing inside or outside. (The sum of these states defines the total probability 
of the particle being in or out.) The total information comprised by the information of all bits on a screen
then represents distances of point particles within (\S3). 

For a detailed encoding, we next consider the problem of elementary memory units, each comprising a certain
fixed number of bits. Relative to the normal to the surface, the notion of a particle being inside or outside applies analogously to 
electric charge, angular momentum and radiation. In a binary encoding, zero electric charge or zero angular momentum would be a result of zero expectation in a binary encoding over all memory units on a screen. A binary encoding for mass, charge and angular momentum as simultaneously measurable quantities comprises 8 states and, correspondingly requires 3 bits. Electromagnetic and gravitational radiation each have two polarizations and their wave vector may point inwards or outwards; each representing 4 states, they combined represent a further 8 states. 

This heuristic discrete counting argument shows a {\em minimal encoding} of 16 different states, which requires memory units of at least 
4 bits (a nibble). 

\section{A covariant formulation}

Our starting point is (\ref{EQN_I4}-\ref{EQN_I4b}) and the encoding of enclosed mass in 1 out of 4 bit memory on the screen.
If $k$ is a local wave number of the de Broglie matter wave and $dr$ is a coordinate distance, then
\begin{eqnarray}
dI = 2\pi d\varphi
\label{EQN_Q2}
\end{eqnarray}
where $d\varphi=kdr$ is a corresponding phase difference of the de Broglie wave as a general expression in curved space-time. 

For a covariant formulation of (\ref{EQN_Q2}) in spherically symmetric space-time, recall the 
Schwarzschild space-time in spherical coordinates $(t,r,\theta,\phi)$. In geometrical units (Newton's constant and 
the velocity of light satisfying $G=c=1$) is described by the line-element
\begin{eqnarray}
ds^2 = - \alpha^2 dt^2 + \frac{1}{\alpha^2} dr^2 + r^2 d\theta^2 + r^2 \sin^2\theta d\phi^2
\label{EQN_Q3}
\end{eqnarray}
with redshift factor $\alpha = \sqrt{1-2M/r}$. Outside the Schwarzschild radius $R_S=2M$, we have
\begin{eqnarray}
k = \frac{k_0}{\alpha},
\label{EQN_Q4}
\end{eqnarray} 
where $k_0=\omega = d\varphi/dt = M$ is the time rate-of-change of phase as measured at infinity. The distance of a spherical screen of radius $r$ to the event horizon is
\begin{eqnarray}
\Delta \varphi = k_0 \int_{R_S}^r \frac{dr}{\alpha} = M [f(x)]_{x=2}^x,  
\label{EQN_Q5}
\end{eqnarray}
where $x=r/M$ and
\begin{eqnarray}
f(x) = x_h+\alpha x+ \ln(\alpha x + x-1),
\label{EQN_Q6}
\end{eqnarray}
where $x_h$ denotes an integration constant. 
Integration of (\ref{EQN_Q2}) thus generalizes (\ref{EQN_I4b}) to
\begin{eqnarray}
A_I = 2\pi l_p^2 \Delta \varphi = \left\{ \begin{array}{cc} 2 \pi Ms_h  & (r=R_S) \\ \\ 2\pi M r & (r>> R_S) \end{array}\right. 
\label{EQN_Q7}
\end{eqnarray}   
where $s_h=Mx_h$ represents the scaled integration constant in (\ref{EQN_Q6}). 

To illustrate our approach of conformal quantization of holographic information in the Einstein areas $A_E$ 
of discretize surface elements, we next turn to black holes with charge. 

\subsection{Reissner-Nordstr\"om black holes}

For a charged black hole, consider the de-redshifted total mass energy $m=m(r)$ within a screen at $r$,
\begin{eqnarray}
m = M - {\cal E}_Q,~~{\cal E}_Q= \frac{Q^2}{2r}.
\label{EQN_Q8}
\end{eqnarray}
By (\ref{EQN_Q7}), the Einstein area $A_{E}^-$ attains the limit $4A_{I}^-=8\pi m s_h$
in a one-bit encoding of the enclosed mass-energy $m$, i.e., when
\begin{eqnarray}
8 \pi m s_h = 4\pi R^2_S.
\label{EQN_9a}
\end{eqnarray}
A two-bit encoding of $M$, comprising $m$ and ${\cal E}_Q$, gives for the total Einstein area per unit surface area $A=4\pi r^2$
\begin{eqnarray}
n \equiv \frac{A_{E}^-+A_{E}^+}{A} = 1 + \frac{Q^2}{R^2_S},
\label{EQN_ndef}
\end{eqnarray}   
where $A_E^+=8\pi {\cal E}_Q r$ by (\ref{EQN_Q7}). Here, the second equation holds at saturation $A_E^-=4\pi R_S^2$, when the Einstein area equals a wave-front area. In this saturated limit, the information density involving two bits is $n/4$. We have, by (\ref{EQN_Q7}) once more,
\begin{eqnarray}
\frac{8 \pi M s_h}{n}= 4\pi  R_S^2.
\label{EQN_9b}
\end{eqnarray}
Combined, the equations (\ref{EQN_9a}-\ref{EQN_9b}) give
\begin{eqnarray}
R_S^2-2Ms_h + Q^2(s_h/R_S)=0,~~R_S^2-2Ms_h+Q^2=0.
\end{eqnarray} 
The solution is the familiar Reissner-Nordstr\"om outer radius
\begin{eqnarray}
R_S = M + \sqrt{M^2-Q^2},~~s_h=R_S.
\label{EQN_RS}
\end{eqnarray}

A convenient parameterization of (\ref{EQN_RS}) is  $Q/M=\sin\lambda$, whereby $R_S = 2M \cos^2(\lambda/2)$.
By (\ref{EQN_RS}), (\ref{EQN_9a}) explicitly expresses the surface area $A_H=4\pi R_S^2$ of the black hole
by the enclosed mass-energy
\begin{eqnarray}
A_H = 16 \pi m^2.
\label{EQN_AHS}
\end{eqnarray}
Upon associating $A_H$ with entropy \cite{bek73,haw74}, 
$A_H$ is non-decreasing by the second law of thermodynamics (e.g. \cite{maj98}), 
whence $m$ is commonly referred to as the irreducible mass. The result is 
\begin{eqnarray}
m = \sqrt{\frac{A_H}{A_S}} M = M \cos^2(\lambda/2),~~{\cal E}_Q=M\sin^2(\lambda/2),
\label{EQN_AA0}
\end{eqnarray}
where $A_S=16\pi M^2$ denotes the horizon surface area of a Schwarzschild black hole of total mass $M$. 
Note that $m=M/2$ for an extremal Reissner-Nordstr\"om black hole associated with $A_H/A_S=1/4$.

\subsection{Extremal Kerr black holes}

Equation (\ref{EQN_9b}) illustrates that $A_H$ decreases with an increase in the free energy in the electromagnetic
field. A similar result holds as a result of energy in angular momentum. Since angular momentum is a vector, the
area decrease is defined locally by the component of total angular momentum normal to each surface element of 
the screen. 
 
According to (\ref{EQN_L12f}), the specific information density can be evaluated at large distances, i.e., on the 
celestial sphere in an asymptotically flat space. Adopting a spherical coordinate system $(r,\theta,\phi)$ alined
with $J$. The above mentioned scaling is hereby maximal at a polar surface element and zero on the equator. 
The Einstein area assumed for angular momentum is hereby less than that for encoding $Q$ by a factor
\begin{eqnarray}
\eta = \frac{\int_0^{\frac{\pi}{2}} \cos\theta \sin\theta d\theta}{\int_0^{\frac{\pi}{2}} \sin\theta d\theta} = \frac{1}{2}.
\label{EQN_eta}
\end{eqnarray}
The reduction in $A_H/A_S$ of an extremal Kerr black hole is hereby more moderate than that for an extremal 
Reissner-Nordstr\"om black hole of the same total mass $M$ by a factor of two, giving $A_H/A_S=1/2$. 
Let ${\cal E}_{J}$ denote the rotational energy associated with $J$. Equation (\ref{EQN_AA0}) now implies
\begin{eqnarray}
m= M-{\cal E}_{J} \ge \frac{1}{\sqrt{2}} M.
\label{EQN_mrot}
\end{eqnarray}
Equality in (\ref{EQN_mrot}) obtains the extreme limit of the general 
Kerr solution ${\cal E}_{J} = 2M \sin^2(\mu/4)$, $J=aM$, $\sin\mu=a/M$, satisfying $m=M\cos(\mu/2)$.  
A generalization of our arguments to deriving this more general result seems of interest, 
but falls outside the scope of the present discussion. 

\section{Computation in cosmology}

For a cosmological distribution of matter, we sum (\ref{EQN_I4}) over the wave numbers $k$ associated with 
each galaxy with $r$ given by the Hubble radius $R_H$, that sets the maximal size of a screen encoding
our visible Universe, giving
\begin{eqnarray}
A_I = 2\pi N R_g R_H,
\label{EQN_I7}
\end{eqnarray}
where $N\simeq 10^{11}$ denotes the number of galaxies in the visible universe, $R_g=GM_g/c^2\simeq 10^{16}$ cm 
is the gravitational radius of the baryonic mass $M_g=10^{11}M_\odot$ of a typical galaxy and 
$R_H=c/H_0\simeq 1.4\times 10^{28}$ cm for a Hubble constant $H_0\simeq 68$ km s$^{-1}$ Mpc$^{-1}$ \cite{pla13}.
Encoding the positions of $M=NM_g$, $I$ is about 20\% of the total information that can be encoded on
the cosmological event horizon, since $A_E/A_H=0.22$, where $A_H=4\pi R_H^2$. 
Including dark matter, $I$ is a fraction of essentially unity of the cosmological information bound on phase space.
The large cosmological phase space is not superfluous, but should be viewed for its potential computational effort \cite{llo02}, and here identified with the encoding of location of all matter in the visible universe. 

Our cosmological information $I$ appears to be particularly large when compared with the total entropy $S$ of the universe. This is remarkable in view of the exceedingly uniform distribution of the cosmic microwave background (and the large scale
structure of matter) in the Universe \cite{ste12}.
Presently, the entropy is about $10^{87}$ in the cosmic microwave background and well over $10^{100}$ in 
supermassive black holes (\cite{ven99,ste12b}, (\ref{EQN_r1}-\ref{EQN_r3}) below), representing 0.1\% of the mass of their host galaxies. These two contributions are exceedingly low compared to the present cosmological entropy bound by a factor of, respectively, about $10^{36}$ and, respectively, $10^{23}$. 
Has the Universe always been in a state of efficient computing?

Following thermodynamic arguments, consider the non-relativistic state and leading order evolution of the universe.
Below a redshift $z_e$ at which the energy density in radiation dropped to that of matter (dark matter and baryonic 
matter), evolution follows a matter dominated universe up to about $z\simeq 0.5$ \cite{rie04,li13}, beyond which there is a 
gradual transition to a de Sitter phase \cite{per99,rie98}. For a leading order estimate, we here follow a matter dominated 
evolution up to the present. Based on Planck \cite{pla13}, we estimate a redshift $z_e\simeq 20,000$ at a temperature 
$T_e\sim 6\times 10^4$ K (cf. \cite{oha94}).

The ratio of entropy in the visible universe is mostly $S=(4/3)\alpha T^3V$ in electromagnetic radiation, where
$\alpha=\pi^2 k_B^2/15c^3\hbar^3= 7.56 \times 10^{-15}$ erg cm$^{-3}$ $K^{-4}$ is the black body constant,
$k_B=1.38\times 10^{-16}$ erg K$^{-1}$ is the Boltzmann constant and $V$ denotes the Hubble volume. 
Expressed as $r=S/S_H$ relative to the maximal entropy $S_H=k_B A_H/4l_p^2$ 
of the cosmological event horizon, we have
\begin{eqnarray}
r=\frac{16\alpha }{9k_B} T^3 R_H l_p^2, 
\label{EQN_r1}
\end{eqnarray}
Thus, the present value $r=r_0$ is $6.7\times 10^{-35}$. Upon including the entropy of supermassive black holes,
$r$ remains small given the limited ratio of $\sim 5\times10^{-17}$ in the total surface area of their
event horizons to that of the cosmological event horizon. 
The ratio (\ref{EQN_r1}) attained a value $r_e$ at $z_e$ higher than the present value by a limited factor
\begin{eqnarray}
\frac{r_e}{r_0}=\left(\frac{T_e}{T_0}\right)^3 \left(\frac{1}{1+z_e}\right)^{3/2}\simeq 2.6\times 10^{6}.
\label{EQN_r2}
\end{eqnarray}
During a prior epoch of inflation, consider the Sitter temperature $k_BT=H\hbar/2\pi$ \cite{unr76,gib77}. 
We then have (adapted from \cite{cai05})
\begin{eqnarray}
k_BT = \frac{H\hbar}{4\pi} \left(1-q\right)\le \frac{H\hbar}{2\pi} 
\end{eqnarray}
in terms of the deceleration parameter $q= -{\ddot{a}a}/{\dot{a}^2}.$
The entropy $S_{0}=(4/3)\alpha T^3$ in the de Sitter radiation hereby satisfies $S_{0}\le 2/135$. Relative to
$S_H\ge 4l_p^2$, we have
\begin{eqnarray}
r \le \frac{1}{270} < 0.37\%.
\label{EQN_r3}
\end{eqnarray}
Combined, (\ref{EQN_r1}), (\ref{EQN_r2}) and (\ref{EQN_r3}) show that $r$ remained small at all times.

\section{Conclusion and discussion}

We derive information in holographic screens from transition probabilities across a cut of the 
quantum mechanical de Broglie wave function.
Two-dimensional surfaces hereby carry Shannon information representing exact unitarity of the binary alternatives 
of particle detections inside or outside. The latter is readily extended to radiation propagating in or out. 
At a minimum, binary encoding of mass, charge, angular momentum or radiation requires surface elements
containing four bits.

Encoding an enclosed (de-redshifted) mass energy $m$ requires an information 
\begin{eqnarray}
I=2\Delta\varphi,~ I=12\Delta\varphi,~~I=2\pi \Delta\varphi
\end{eqnarray}
at a distance measured in de Broglie phase difference $\Delta\varphi$ away from a flat screen, at the center of a cube of linear size $2\Delta\varphi$ and at the center of a spherical screen of radius $\Delta\varphi$, respectively. 
For the first, discretization in bits is closely associated
with the Heisenberg uncertainty relation (\ref{EQN_HE}). 
For the latter, an extremal cut obtains in the limit as the Einstein area $A_E=4I$ equals the wave front 
area $A_f$, i.e., a minimal screen enveloping the event horizon of a black hole in the encoding of mass enclosed
within by four-bit surface elements. In free fall, the screen would become null and its information would
appear as entropy to an outside observer. For Reissner-Nordstr\"om black holes, a reduced area of the event horizon 
represents the mass-encoding of irreducible mass within, given by the total mass-energy less the mass-energy in 
the external electromagnetic field. Encoding specific to the type of mass-energy, see further \cite{wal84,epp77,fla91}, 
explains the apparent discrepancy by a factor of two in de Hoop conjecture \cite{fla91} arising from 
electric charge \cite{bon83,pon87}. Similar considerations seem to apply to black holes with angular momentum.

In deriving information from a propagator, there is no local coupling in emergent gravitation (e.g. \cite{ver11}) to
a four-covariant vacuum. For instance, the cosmological constant $\Lambda_0\simeq l_p^{-4}$
\cite{wei89} predicted by quantum field theory gives rise to a four-covariant stress-energy tensor $\Lambda_0g_{ab}$. 
The associated propagator is trivial, given a zero prior on the number of particles. The information 
associated with any cut (any screen) is hereby zero, since there is nothing to image.
This decoupling of $\Lambda_0$ obviates the need for the Universe to be finite \cite{kal14}. 
The small positive value of the observed cosmological constant $\Lambda$ inferred from the present epoch 
of accelerated expansion \cite{rie04} has a different origin, e.g., in a finite de Sitter 
temperature of the cosmological event horizon \cite{eas11}. 

According (\ref{EQN_I7}), our cosmological information $I$ representing the matter distribution in the visible Universe is about maximal. The latter has previously been considered as a bound on the rate of computation in the Universe \cite{llo02}. 
Here, we identify this bound with the information required in imaging the microphysical distribution of matter in the Universe. Present concordance with a near-maximum deepens the already mysteriously low 
entropy in the universe. Since the answer is not explained by basic thermodynamic considerations (\S8), 
it appears to be related to the problem of extremely efficient computing in Nature.

Following \cite{mar98,llo02}, a given mass energy $M_0c^2$ defines a rate of computation set by the rate of change
in total phase in units of $\pi/2$. (One pass through $2\pi$ is a four bit calculation.) The present cosmological computational effort in imaging the visible 
Universe is on the order of
\begin{eqnarray}
\dot{n}_0=\frac{2M_0c^2}{\pi \hbar}=\frac{c^5}{H_0 G \pi \hbar}\simeq 5\times 10^{103} ~\mbox{s}^{-1},
\label{EQN_dN}
\end{eqnarray}
where $M_0=NM_g$ by (\ref{EQN_I7}). 
We can extrapolate to the future based on the currently observed transition to the accelerated
expansion state of the Universe, characterized by $q_0<0$ \cite{rie04} with a positive Hubble flow of galaxies 
passing through the cosmological event horizon. This Hubble flow implies a gradual decrease in $A_I$ 
as information leaves the visible Universe. Using the present Hubble constant as a proxy for the future in 
a standard $\Lambda$CDM cosmology, $M(t)\simeq M_0 e^{-3H_0 t}$. By (\ref{EQN_dN}), the future of the
visible Universe holds a finite number of computations
\begin{eqnarray}
n\simeq \frac{\dot{n}_0}{3H_0} \simeq  8\times 10^{120}.
\end{eqnarray}
At large time in the future dark energy appears as an entropic remnant of the latter on a Hubble time scale (cf. \cite{per09})
and calculations gradually cease. 

{\bf Acknowledgments.} The author thanks Ee Chang-Young and Paul Steinhardt for stimulating discussions,
and constructive comments from the anonymous referee.

\end{document}